\documentclass[twocolumn]{aastex631}

\usepackage{xspace}

\usepackage{amsmath}
\usepackage{natbib}
\usepackage{enumitem}
\usepackage[caption=false]{subfig}
\usepackage{graphicx}
\usepackage[T1]{fontenc}
\usepackage{booktabs}
\usepackage[encapsulated]{CJK}
\usepackage[utf8]{inputenc}
\usepackage{subfig}

\newcommand{\halpha}{\ensuremath{\textrm{H}\alpha}\xspace}

\newcommand{\hbeta}{\ensuremath{\textrm{H}\beta}\xspace}
\newcommand{\pabeta}{\ensuremath{\textrm{Pa}\beta}\xspace}

\newcommand{\NII}{\ensuremath{[\mathrm{N}\textsc{ ii}]}\xspace}

\newcommand{\FeII}{\ensuremath{[\mathrm{Fe}\textsc{ ii}]}\xspace}

\newcommand{\AV}{\ensuremath{A_{\mathrm{V}}}\xspace}
\newcommand{\Aha}{\ensuremath{A_{\mathrm{\halpha}}}\xspace}
\newcommand{\Ahaneb}{\ensuremath{A_{\mathrm{\halpha, \mathrm{neb}}}}\xspace}

\newcommand{\F}
{\ensuremath{F}\xspace}

\newcommand\ngals{66 }  
\newcommand\npossible{766 } 
\newcommand\nmosdef{396 } 
\newcommand\nhaonly{143 } 
\newcommand\ncombined{209 }

\shorttitle{Dust Geometry at $z\sim2$ from MegaScience Medium Bands}
\shortauthors{Lorenz et al.}

\graphicspath{./}

\begin{document}

\title{Evidence for Shallow Nebular Attenuation Curves and Patchy Dust Geometry at $\boldsymbol{z\sim2}$ with \pabeta/\halpha Measurements from JWST-MegaScience Medium Band Photometry}

\author[0000-0002-5337-5856]{Brian Lorenz}
\affiliation{Department of Astronomy, University of California, Berkeley, CA 94720, USA}

\author[0000-0002-1714-1905]{Katherine A. Suess}
\affiliation{Department for Astrophysical and Planetary Science, University of Colorado, Boulder, CO 80309, USA}

\author[0000-0002-7613-9872]{Mariska Kriek}
\affiliation{Leiden Observatory, Leiden University, P.O. Box 9513, 2300 RA Leiden, The Netherlands}

\author[0000-0002-0108-4176]{Sedona H. Price}
\affiliation{Space Telescope Science Institute, 3700 San Martin Drive, Baltimore, Maryland 21218, USA}

\author[0000-0001-6755-1315]{Joel Leja}
\affiliation{Department of Astronomy \& Astrophysics, The Pennsylvania State University, University Park, PA 16802, USA}
\affiliation{Institute for Computational \& Data Sciences, The Pennsylvania State University, University Park, PA 16802, USA}
\affiliation{Institute for Gravitation and the Cosmos, The Pennsylvania State University, University Park, PA 16802, USA}

\author[0000-0002-7570-0824]{Hakim Atek}
\affiliation{Institut d’Astrophysique de Paris, CNRS, Sorbonne Universit\'e, 98bis Boulevard Arago, 75014, Paris, France}

\author{Abhiyan Barailee}
\affiliation{Department for Astrophysical and Planetary Science, University of Colorado, Boulder, CO 80309, USA}

\author[0000-0001-5063-8254]{Rachel Bezanson}
\affiliation{Department of Physics and Astronomy and PITT PACC, University of Pittsburgh, Pittsburgh, PA 15260, USA}

\author[0000-0001-6755-1315]{Gabriel Brammer}
\affiliation{Cosmic Dawn Center (DAWN), Copenhagen, Denmark}
\affiliation{Niels Bohr Institute, University of Copenhagen, Jagtvej 128, Copenhagen, Denmark}
\author[0000-0002-7031-2865]{Sam E. Cutler}
\affiliation{Department of Astronomy, University of Massachusetts, Amherst, MA 01003, USA}

\author[0000-0001-8460-1564]{Pratika Dayal}
\affiliation{Kapteyn Astronomical Institute, University of Groningen, 9700 AV Groningen, The Netherlands}

\author[0000-0002-2380-9801]{Anna de Graaff}
\affiliation{Max-Planck-Institut f\"ur Astronomie, K\"onigstuhl 17, D-69117, Heidelberg, Germany}

\author[0000-0002-5612-3427]{Jenny E. Greene}
\affiliation{Department of Astrophysical Sciences, Princeton University, 4 Ivy Ln., Princeton, NJ 08544, USA}

\author[0000-0001-6278-032X]{Lukas J. Furtak}
\affiliation{Physics Department, Ben-Gurion University of the Negev, P.O. Box 653, Be’er-Sheva 84105, Israel}

\author[0000-0002-2057-5376]{Ivo Labb\'e}
\affiliation{Centre for Astrophysics and Supercomputing, Swinburne University of Technology, Melbourne, VIC 3122, Australia}

\author[0000-0001-9002-3502]{Danilo Marchesini}
\affiliation{Department of Physics \& Astronomy, Tufts University, MA 02155, USA}

\author[0000-0003-0695-4414]{Michael V. Maseda}
\affiliation{Department of Astronomy, University of Wisconsin-Madison, 475 N. Charter St., Madison, WI 53706, USA}

\author[0000-0001-8367-6265]{Tim B. Miller}
\affiliation{Center for Interdisciplinary Exploration and Research in Astrophysics (CIERA), Northwestern University, 1800 Sherman Ave, Evanston IL 60201, USA}

\author[0000-0002-9816-9300]{Abby Mintz}
\affiliation{Department of Astrophysical Sciences, Princeton University, 4 Ivy Ln., Princeton, NJ 08544, USA}

\author[0000-0001-7300-9450]{Ikki Mitsuhashi}
\affiliation{Department for Astrophysical and Planetary Science, University of Colorado, Boulder, CO 80309, USA}

\author[0000-0003-2804-0648]{Themiya Nanayakkara}
\affiliation{Centre for Astrophysics and Supercomputing, Swinburne University of Technology, P.O. Box 218, Hawthorn VIC 3122, Melbourne, Australia}
\affiliation{JWST Australia Data Centre, Swinburne Advanced Manufacturing and Design Centre, John Street, Hawthorn VIC 3122, Australia}

\author[0000-0002-7524-374X]{Erica Nelson}
\affiliation{Department for Astrophysical and Planetary Science, University of Colorado, Boulder, CO 80309, USA}

\author[0000-0002-9651-5716]{Richard Pan}
\affiliation{Department of Physics \& Astronomy, Tufts University, MA 02155, USA}

\author[0009-0001-0715-7209]{Natalia Porraz Barrera}
\affiliation{Department for Astrophysical and Planetary Science, University of Colorado, Boulder, CO 80309, USA}

\author[0000-0001-9269-5046]{Bingjie Wang (\begin{CJK*}{UTF8}{gbsn}王冰洁\ignorespacesafterend\end{CJK*})}
\affiliation{Department of Astronomy \& Astrophysics, The Pennsylvania State University, University Park, PA 16802, USA}
\affiliation{Institute for Computational \& Data Sciences, The Pennsylvania State University, University Park, PA 16802, USA}
\affiliation{Institute for Gravitation and the Cosmos, The Pennsylvania State University, University Park, PA 16802, USA}

\author[0000-0003-1614-196X]{John R. Weaver}
\affiliation{Department of Astronomy, University of Massachusetts, Amherst, MA 01003, USA}

\author[0000-0003-2919-7495]{Christina C.\ Williams}
\affiliation{NSF’s National Optical-Infrared Astronomy Research Laboratory, 950 North Cherry Avenue, Tucson, AZ 85719, USA}

\author[0000-0001-7160-3632]{Katherine E. Whitaker}
\affiliation{Department of Astronomy, University of Massachusetts, Amherst, MA 01003, USA}
\affiliation{Cosmic Dawn Center (DAWN), Copenhagen, Denmark}

\begin{abstract}
We constrain the nebular attenuation curve and investigate dust geometry in star-forming galaxies at cosmic noon using photometric medium-band emission line measurements. We measure \halpha emission line fluxes for a sample of \ncombined star-forming galaxies at $1.2<z<2.4$ in MegaScience/UNCOVER with stellar masses spanning $7.85<\log_{10}(M_*/M_\odot)<11.0$. For \ngals of these galaxies, we also measure a \pabeta flux. We find that the \pabeta/\halpha line ratio increases strongly with stellar mass and star-formation rate (SFR) across our full mass range, indicating that more massive galaxies are dustier. We compare our results with a mass-, SFR-, and redshift-matched sample of galaxies from the MOSDEF survey with spectroscopic measurements of \halpha/\hbeta, finding that a shallow \citet{reddy_jwstaurora_2025} nebular attenuation curve is more consistent with our observations than the typically assumed \citet{cardelli_relationship_1989} attenuation curve, especially for massive galaxies. This shallow attenuation curve could be explained by low dust covering fractions in star-forming regions. Through comparison to other studies, we show that assuming this shallower attenuation curve can increase the inferred \Ahaneb by up to 1 magnitude at high masses. We observe no trend between \Ahaneb and axis ratio, indicating that nebular attenuation is likely localized to small clumps. Altogether, our results strongly suggest that dust geometry is patchy and non-uniform, especially in massive galaxies. Our results highlight the ability of JWST medium bands to probe emission lines for large samples of galaxies, and statistically constrain dust properties in upcoming large programs.
\end{abstract}


\section{Introduction} \label{sec:intro}

The distribution of dust within a galaxy affects the observed attenuation, and therefore the observed flux. If the dust is not modeled correctly, it dramatically alters the measured mass, star-formation rate (SFR), and emission line properties \citep{salim_dust_2020}. Therefore, properly understanding dust content and geometry is essential for accurate measurements of galaxy properties, especially for systems with complicated dust distributions. Additionally, dust attenuation seems to correlate with stellar mass, SFR, and metallicity \citep[e.g.,][]{garn_predicting_2010, dominguez_dust_2013, price_direct_2014, whitaker_constant_2017, cullen_vandels_2018, battisti_average_2022, runco_mosdef_2022, shapley_mosfire_2022, lorenz_updated_2023, shapley_jwstnirspec_2023, lorenz_stacking_2024, maheson_unravelling_2024, maheson_big_2025}, but the exact relationship between dust and galaxy properties is not fully understood.

At cosmic noon, the study of dust is particularly challenging. It is difficult to obtain high enough spatial resolution to directly observe the location of dust. Instead, we often rely on imperfect measurements, such as SED fitting, IR emission, and emission line ratios. In particular, line ratios are an excellent probe of dust, as they directly measure the relative nebular attenuation between two wavelengths. However, converting this measurement into the attenuation at a specific wavelength (for example, the attenuation at the wavelength of \halpha, \Aha) depends on the shape of the nebular attenuation curve, which is not well constrained. While the \citet{cardelli_relationship_1989} attenuation curve is typically assumed, it does not seem to apply universally to all galaxies, and may be significantly too steep at cosmic noon \citep{reddy_jwstaurora_2025}. 

Determining the correct attenuation curve is extremely important, as a variety of measured properties depend on the dust content. For example, the shape of the nebular attenuation curve affects measurements of SFR derived from Hydrogen emission lines. Similarly, the assumed stellar attenuation curve can dramatically affect galaxy mass and SFR measured from SED fitting \citep{salim_dust_2020}. Additionally, age, dust, and metallicity all affect an SED in similar ways, leading to a degeneracy between the three parameters \citep[e.g.][]{papovich_stellar_2001, tacchella_stellar_2022}.

It is equally important to understand how dust is distributed within a galaxy, since this distribution affects the attenuation curve. Star-forming galaxies at cosmic noon often have complex, clumpy morphologies \citep{elmegreen_resolved_2007, swinbank_hubble_2010, tadaki_nature_2014}. For example, \citet{gillman_structure_2024} find that dusty sub-millimeter galaxies at cosmic noon exhibit a wide variety of morphologies, and \citet{price_uncover_2025} find that their dust lanes trace the stellar continuum. Other studies propose that dust is clumpy at $z\sim2$, with attenuation occurring in star-forming regions rather than the ISM \citep{reddy_mosdef_2015, lorenz_updated_2023, lorenz_stacking_2024, reddy_jwstaurora_2025}, as well as is in large, kiloparsec-scale star-forming clumps \citep{schreiber_constraints_2011, wuyts_smoother_2012}. Efforts have also been made to map out the locations of dust through emission line maps \citep{nelson_where_2016, tacchella_dust_2018, matharu_first_2023, lorenz_measuring_2025}. Taken together, all of the data seem to suggest complicated and clumpy dust distributions. Further studies are necessary to form a more comprehensive picture of dust at cosmic noon.

Detecting multiple hydrogen emission lines --- not just \halpha and \hbeta --- allows us to mitigate uncertainties from the assumed attenuation curve, making it a promising route to further understand dust in distant galaxies. 
Assuming Case B recombination, the line ratios are determined by quantum mechanics, and any deviations in these ratios are due to dust \citep[e.g.,][]{reddy_paschen-line_2023}. Typically, measuring these emission lines requires spectroscopy, which tends to be observationally expensive. Fortunately, studies have been able to measure emission lines directly from photometry, allowing for far larger samples of emission line measurements \citep[e.g.,][]{geach_hizels_2008, kriek_dust_2013, roberts-borsani_improving_2021, simmonds_ionizing_2023, withers_spectroscopy_2023, lorenz_measuring_2025}. In particular, with the launch of the James Webb Space Telescope \citep[JWST,][]{gardner_james_2023} the medium bands are a very promising tool to observe emission line fluxes at cosmic noon \citep{lorenz_measuring_2025}. These measurements have been verified against spectra, and have been shown to be accurate down to line equivalent widths of 10\AA. We use these capabilities to measure \halpha and \pabeta emission lines, whose ratio is sensitive to nebular dust attenuation. Additionally, this photometric method has an advantage of observing entire galaxies in the aperture, avoiding slit losses that come with spectroscopy. While integral field units (IFUs) can also avoid slit losses, the purely photometric technique is much more observationally efficient.

In this work, we measure \halpha emission line fluxes for a sample of \ncombined galaxies at $1.2<z<2.4$, and \pabeta fluxes for \ngals, with stellar masses ranging from $7.85<\log_{10}(M_*/M_\odot)<11.0$ from the UNCOVER/MegaScience survey \citep{bezanson_jwst_2024, suess_medium_2024}, including lower stellar masses than typically observed at this epoch. We use the \pabeta/\halpha ratio to examine the relationship between dust, mass, and SFR. Then, we match our data with the MOSDEF survey \citep{kriek_mosfire_2015} on mass, SFR, and redshift, to create a comparable sample of galaxies with \halpha/\hbeta (Balmer decrement) measurements. With multiple line ratios, we place constraints on the nebular attenuation curve. Finally, we investigate the relationship between \Ahaneb and galaxy inclination, drawn from \texttt{Pysersic} fits to the photometry. We discuss the implications for dust geometry based on all of our findings.

Throughout this work we assume a $\Lambda$CDM cosmology with $\Omega_m=0.3$, $\Omega_{\Lambda}=0.7$, and $H_0=70$ km s$^{-1}$ Mpc$^{-1}$. 

\section{Data and Sample Selection} \label{sec:data}

\subsection{MegaScience Photometric Sample} \label{subsec:megascience_sample}

The UNCOVER and MegaScience surveys have obtained NIRCam photometric coverage of the Abell 2744 cluster, a wide gravitational lens with high magnification \citep{lotz_frontier_2017}. UNCOVER observed broad-band photometry for over 70,000 objects \citep{bezanson_jwst_2024}. MegaScience supplemented UNCOVER by filling in all of the JWST medium bands over the same area, dramatically increasing the sampling of the measured spectral energy distributions (SEDs) for galaxies in the field \citep{suess_medium_2024}. Together, these surveys provide precise photometry for a large number of objects at $1.2 < z< 2.4$ where \halpha and \pabeta are contained within medium bands. 

For this work, we use the public DR3 photometric catalog \citep{suess_medium_2024}. All of the $\sim$70,000 sources are fit with \texttt{Prospector}-$\beta$ priors on mass and star formation history and self-consistent lens modeling \citep[][v5.3.0]{wang_uncover_2024}. Through exploring the parameter space with \texttt{dynesty} \citep{speagle_dynesty_2020}, these fits provide photometric redshifts, stellar mass, SFR, metallicity, and dust measurements for every object. We also use the morphology catalog (Zhang et al. in prep), which contains \texttt{Pysersic} fits \citep{pasha_pysersic_2023} to the photometry in each band. Each object is fit with a single sersic profile in all bands where the galaxy is detected with a signal-to-noise ratio (SNR) of at least 10. Each band is fit independently using an observed PSF built from bright point sources in the field (see \citet{suess_medium_2024, weaver_uncover_2024}). Sources with SNR$>5$ within 2~arcsec and within 2 magnitudes of the primary source are fit simultaneously, while other sources are masked. Particularly useful for this work, the morphology catalog includes axis ratio measurements. 

\begin{figure}[]
\vglue -5pt
\includegraphics[width=0.5\textwidth]{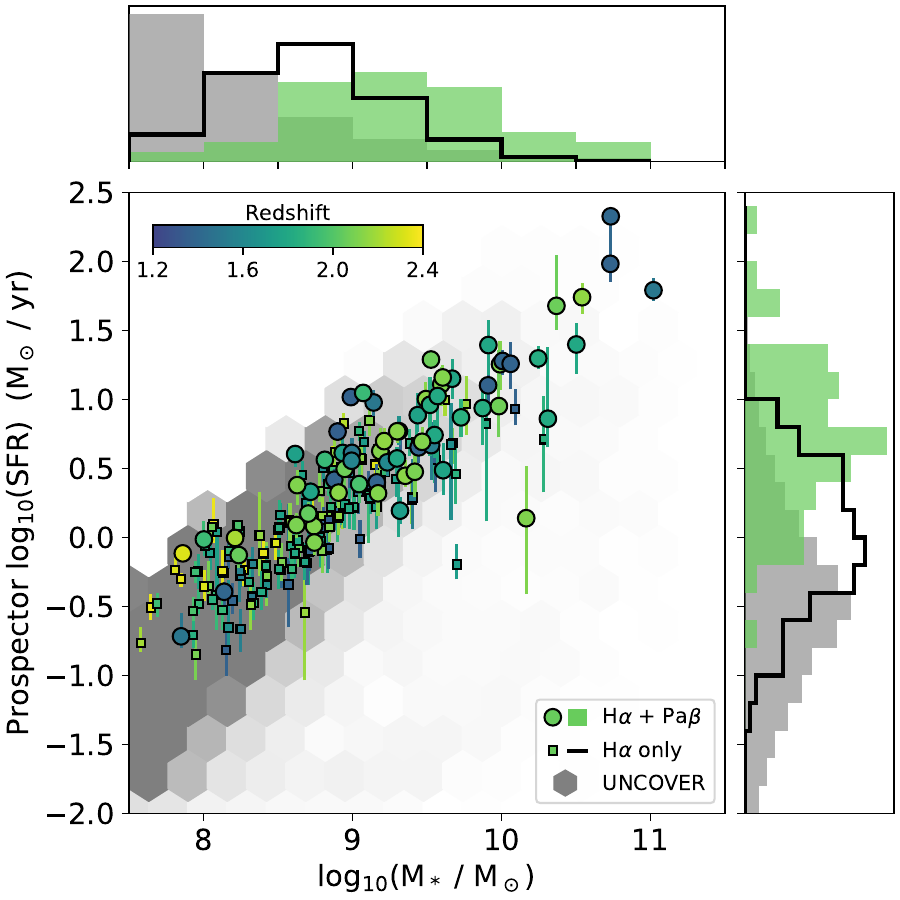}
\caption{
\texttt{Prospector} SFR vs stellar mass for the \ngals galaxies in our sample with both \halpha and \pabeta measurements (circles, colored by redshift, green histogram), the \nhaonly galaxies with \halpha detections and \pabeta upper limits (squares, black histogram), and the full UNCOVER sample (gray hexagons, gray histogram). Histograms for SFR and stellar mass for the three populations are also shown. Our sample captures galaxies down to $\log_{10}(M_*/M_\odot) = 7.85$. They span the full range of redshifts from $1.2<z<2.4$, and do not exhibit mass or SFR trends with redshift.  
}
\label{fig:sample_select}
\end{figure}

We select a sample of galaxies for which we can measure \halpha and \pabeta emission line strengths from medium-band imaging, as demonstrated by \citet{lorenz_measuring_2025}. First, we ensure that each emission line falls within a medium band, and that there are surrounding medium bands that can be used for continuum measurement. For each object, we use the \texttt{Prospector} photometric redshift distribution to ensure that the 5-95\% confidence interval places both \halpha and \pabeta emission lines within a medium band. This selection leaves \npossible galaxies that could potentially have these emission lines measured through the medium bands. However, we further restrict our sample to ensure that the line measurements are of high enough quality to mimic JWST Prism spectroscopic measurements, following similar restrictions to \citet{lorenz_measuring_2025}. Objects close to the bright cluster galaxies (BCGs) may be contaminated by their light, so we remove any objects with modeled BCG surface brightness higher than 0.4 (nJy / pix$^2$) (104 galaxies). Objects above this threshold showed contamination from a visual inspection. Then, we require a signal-to-noise ratio of at least 3 in \halpha flux (285 galaxies) (see Section \ref{sec:data_analysis}). Finally, some galaxies have a poor continuum measurement, typically due to high uncertainties in one of the continuum filters. In these cases, the slope for the continuum is inconsistent with the slope predicted by \texttt{Prospector}. We remove galaxies with significant continuum slope offsets which resulted in anomalously large or small dust measurements (168 galaxies). Our final selected sample of \ncombined galaxies can be seen in Figure \ref{fig:sample_select}. 

Of the \ncombined galaxies in our sample, \nhaonly have \pabeta SNR less than 3, leaving \ngals with detections in both lines. These \pabeta nondetections are preferentially lower mass galaxies, so omitting them would introduce a strong source of bias into our sample. For these systems, we compute an upper limit on their \pabeta line measurement (see Section \ref{sec:data_analysis}). Then, as we carry out our analysis, we measure an upper limit on the dust attenuation. 

While these selection methods ensure high-quality measurements, we note that they may introduce biases. Although we attempt to include many low-mass galaxies as upper limits, the sample is still biased towards higher mass galaxies, as these tend to have more certain photometric redshifts and higher SNR emission lines. The sample may also miss quiescent sources due to low emission line strengths, as well as extremely dusty sources; including these sources could substantially decrease average \halpha fluxes. Out of all possible targets in the UNCOVER survey with \halpha and \pabeta in medium bands, the sample contains 61\% of sources within 0.5\,dex of the \citet{popesso_main_2023} star-forming main sequence down to $\log_{10}(M_*/M_\odot) = 9.5$, and 41\% down to $\log_{10}(M_*/M_\odot) = 9$ (apparent magnitude $\sim23$). We are able to measure a number of fainter galaxies, even down to apparent magnitudes of 26, if they have significantly high equivalent widths for their \halpha and \pabeta emission lines. Despite these biases, our sample is representative of star-forming galaxies above $\log_{10}(M_*/M_\odot) = 9$ with strong emission lines, as it contains a variety of galaxies both above and below the star-forming main sequence.

\begin{figure*}[tp]
\vglue -5pt
\includegraphics[width=\textwidth]{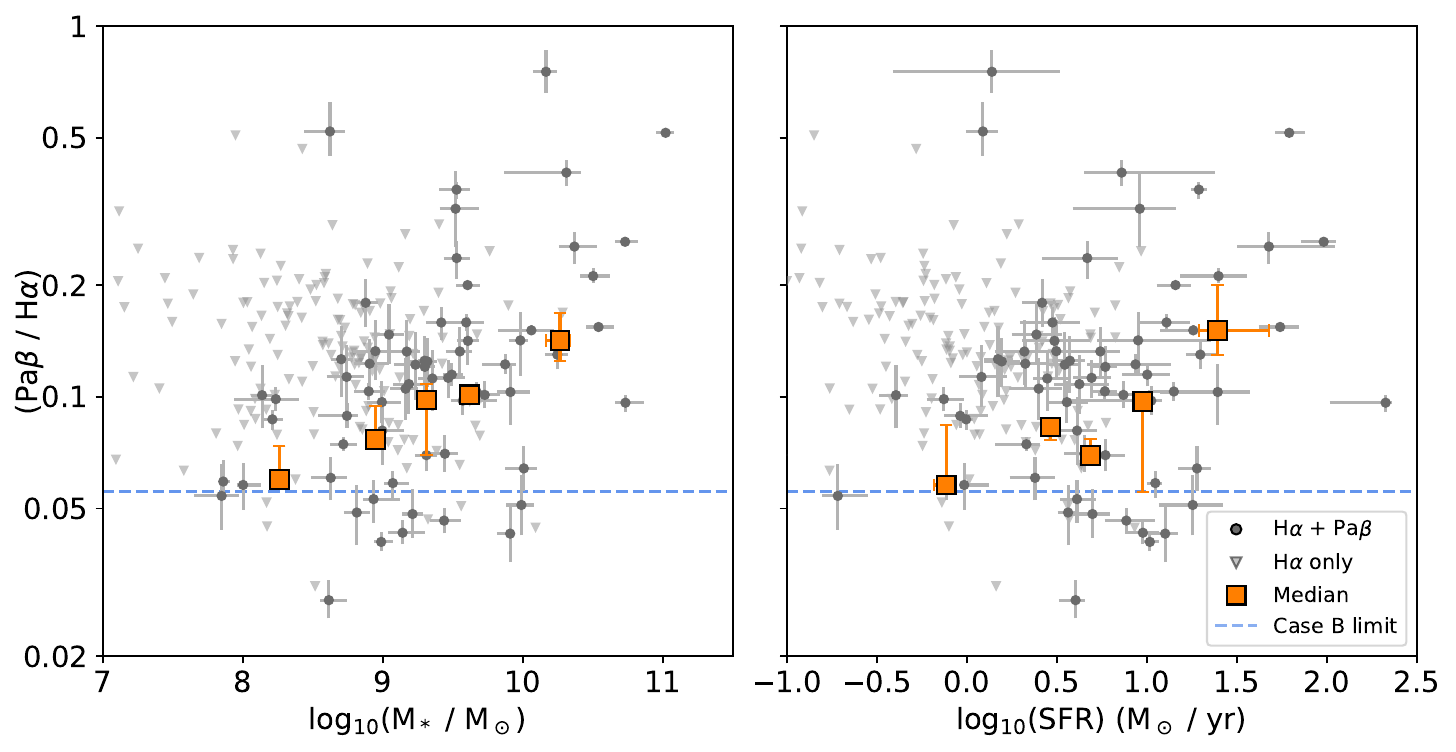}
\caption{Measured \pabeta/\halpha line ratio plotted against stellar mass (left) and SFR (right). A larger line ratio implies a more dusty galaxy, although the exact conversion depends on the assumed attenuation curve. The intrinsic \pabeta/\halpha ratio assuming Case B recombination is 1/18 (0.056). Medians and uncertainties (orange) are computed in bins with equal numbers of galaxies from the fully-detected sample using the \citet{turnbull_empirical_1976} estimator that accounts for the upper limits in each bin. As expected, we see a clear trend of increasing dust attenuation with increasing mass and increasing SFR.}
\label{fig:dust_mass_sfr}
\end{figure*}

\subsection{MOSDEF Comparison Sample}
\label{subsec:mosdef_sample}

In this work, we also make use of a comparable sample of galaxies from the MOSDEF survey \citep{kriek_mosfire_2015} that have spectroscopic \halpha and \hbeta emission line measurements. MOSDEF used the MOSFIRE spectrograph \citep{mclean_mosfire_2012} on the Keck I telescope to take deep optical spectra of $\sim1500$ H-band selected galaxies. The survey covers three redshift bands: $1.37\leq z\leq 1.70$, $2.09\leq z\leq 2.61,$ and $2.95\leq z \leq 3.60$. To roughly match our MegaScience data ($1.3 < z <2.4$), we limit the redshift of the MOSDEF sample to the two lower bands $1.37 < z < 2.61$. Objects were selected from the MOSDEF emission line catalogs \citep{kriek_mosfire_2015, reddy_mosdef_2015}, which contain the absorption-corrected \halpha and \hbeta spectroscopic fluxes. The MOSDEF galaxies have stellar population properties from FAST \citep{kriek_ultra-deep_2009} fits to their SEDs, as well as spectroscopic redshift measurements. 

We compare the MOSDEF galaxies to MegaScience above $\log_{10}(M_*/M_\odot) = 9$, since MOSDEF does not contain measurements for lower mass galaxies. For this mass range, we find that the MOSDEF and MegaScience samples have similar median mass and SFR, though their mean redshifts differ slightly (Table \ref{tab:compare}). 

\begin{table}
\centering
\caption{\label{tab:compare} Average mass, SFR, and redshift for the MegaScience data in this work with $\log_{10}(M_*/M_\odot) > 9$ and the MOSDEF sample. Minima and maxima for each property are shown in small font.}
\begin{tabular}{|c|c|c|c|c|}
\hline
& n & $\log_{10}(M_*)$ & $\log_{10}$(SFR)  & $z$ \\
& &$M_\odot$&$M_\odot$yr$^{-1}$&\\
\hline
MegaScience & 43 & $9.68_{9.04}^{10.73}$ & $1.25_{0.14}^{2.33}$ & $1.83_{1.36}^{2.15}$ \\
\hline
MOSDEF & 382 & $9.93_{9.05}^{10.79}$ & $1.05_{-0.09}^{2.17}$  & $2.06_{1.24}^{2.58}$  \\
\hline
\end{tabular}
\end{table}

We also remove active galactic nuclei (AGN) from the MOSDEF survey, as emission line contributions from AGN can easily affect the Balmer decrement. We follow the AGN removal process of \citet{sanders_mosdef_2021}, which includes X-ray and IR identification as well as requiring log(\NII/\halpha) $\leq-0.3$. We also require an SNR of 3 for both of the \halpha and \hbeta emission lines, similar to our requirements for MegaScience. From this sample, we take the absorption line corrected \halpha and \hbeta measurements to compute a Balmer decrement for each object. 

In the resulting MOSDEF sample, we have \nmosdef galaxies with a median mass of $\log_{10}(M_*/M_\odot) = 9.91$ and median SFR of $11 M_\odot$yr$^{-1}$. The MOSDEF data also share similar biases to the MegaScience data. Objects with low SFR are likely to have low SNR \hbeta lines, and therefore may drop out of the sample \citep[e.g.,][]{lorenz_updated_2023}. Similarly, extremely dusty objects would show very little \hbeta emission. Due to these similar biases, the MOSDEF and MegaScience samples are both probing similar populations of star-forming galaxies at cosmic noon. 

\section{Emission Line and Dust Measurements} \label{sec:data_analysis}

To derive fluxes from the medium-band photometry, we follow a similar process to \citet{lorenz_measuring_2025}. For each galaxy, we identify the two filters containing \halpha and \pabeta. For each line, we take two continuum filters as the nearest medium bands on either side of the line filter. Instead of using the continuum points directly (as in \citet{lorenz_measuring_2025}), we use a scaled continuum model from \texttt{Prospector}. To build this model, we take the maximum a posteriori (MAP) parameters from the UNCOVER \texttt{Prospector} fits \citep{wang_uncover_2024} and recalculate the spectra without nebular emission. We integrate this model in each of the continuum filters then scale it to match the average flux of the continuum observations. We verify that the slope of the \texttt{Prospector} model matches the slope of the observations with a scaled $\chi^2$ metric (see Section \ref{subsec:megascience_sample}). Finally, we integrate the scaled model over the emission line filter to generate a continuum point that includes stellar absorption in each hydrogen transition. We then carry out the method from \citet{lorenz_measuring_2025} with this new continuum point.

We also correct for contaminating emission lines in each filter similar to \citet{lorenz_measuring_2025}. For the \halpha measurements, we remove \NII contributions based on the fundamental metallicity relation (FMR) from \citet{sanders_mosdef_2021}, using the \texttt{Prospector} SFR and stellar mass measurements. We assume an intrinsic theoretical transition probability ratio of \NII6550\AA\ to \NII6585\AA\ as 1/3 \citep[see][table 1]{dojcinovic_flux_2023}. For \pabeta, we correct for \FeII using the median correction found from all spectroscopic measurements in UNCOVER, as in \citet{lorenz_measuring_2025}. We assume this median \FeII/\pabeta ratio of 0.27 for all objects. This correction introduces a small uncertainty on the \pabeta line flux of roughly 5\%.

Uncertainties on both emission line fluxes are measured through Monte Carlo simulation. We repeat the entire line measurement process, perturbing by known uncertainties on the three photometric points. We take the 16th and 84th percentiles of the resulting distribution as our upper and lower error bars. The uncertainty is used to compute SNR for each line for all objects with detectable \halpha and \pabeta, which we use to determine our sample selection (see Section \ref{subsec:megascience_sample}). For the objects with \pabeta SNR$<3$, we compute $2\sigma$ upper limits on their \pabeta fluxes from their uncertainties. 

We compute median \halpha/\pabeta line ratio measurements in bins of mass and axis ratio. However, the large number of upper-limit data points make the traditional median computation method inaccurate. Instead, we treat the \pabeta non-detections as interval-censored data, with a $2\sigma$ upper limit and a lower limit set by Case B recombination (1/18). While there are some examples of non-Case B recombination observed with JWST in low-mass systems \citep[e.g.,][]{scarlata_universal_2024}, they appear to be rare. There is no evidence yet that these are widespread cases among any given galaxy population. In this work, only 6 of the upper limits fall below the Case B limit, which we treat as uncensored data. Medians with uncertainties can be robustly estimated on interval-censored data by the \citet{turnbull_empirical_1976} estimator, so we apply this technique to all of the medians shown in this work (Figures \ref{fig:dust_mass_sfr}, \ref{fig:attenuation_curve}, \ref{fig:comparison_to_others}, \ref{fig:dust_axisratio}). Uncertainties on the median are computed by bootstrapping. Additionally, we also ran the analysis without including the Case B limit on the \pabeta nondetections, allowing them to be unbounded upper limits. The medians do not change substantially, and the conclusions remain the same.

\section{Results}
\label{sec:results}

We now have a dataset of \ngals galaxies with measured \halpha and \pabeta emission lines, nebular \Aha, \texttt{Prospector} mass and SFR, and \texttt{Pysersic} morphological parameters. We have an additional \nhaonly galaxies with \halpha measurements and \pabeta upper limits, leading to upper limits on \Ahaneb. Our data extend to 0.5\,dex lower in stellar mass than most previous dust analysis in this redshift regime, reaching a few galaxies with $\log_{10}(M_*/M_\odot) < 8$. In this section, we correlate the measured \Ahaneb values with mass and SFR. Then, through comparison to MOSDEF data that includes \halpha and \hbeta emission lines, we constrain the nebular attenuation curve. Finally, we make inferences about dust geometry with additional correlations between nebular \Aha and axis ratio.

\subsection{Dust, Mass, and SFR}

Our large sample of \pabeta/\halpha line ratio measurements allow us to evaluate dust properties of galaxies at cosmic noon. We assess trends between line ratio and mass, as well as line ratio and SFR in Figure \ref{fig:dust_mass_sfr}. We show both individual points as well as five median bins. Each median bin contains an equal number of galaxies from the sample with both \halpha and \pabeta detected. The median values and uncertainties are computed from bootstrapping the \citet{turnbull_empirical_1976} estimator, including both the measurements and upper limits. There is a clear trend of increasing \pabeta/\halpha line ratio, and therefore inferred \Ahaneb, as stellar mass increases. Many prior works also find that nebular attenuation strongly correlates with stellar mass \citep[e.g.,][]{garn_predicting_2010, dominguez_dust_2013, price_direct_2014, whitaker_constant_2017, cullen_vandels_2018, battisti_average_2022, shapley_mosfire_2022, runco_mosdef_2022, maheson_unravelling_2024, lorenz_stacking_2024}. Similarly, \pabeta/\halpha line ratio shows a strong trend with increasing SFR. In summary, galaxies with higher mass and SFR are more dusty.

\subsection{Nebular Attenuation Curves}
\label{subsec:SED_dust}

In order to constrain the nebular attenuation curve, we require at least three hydrogen emission lines at different wavelengths. Then, assuming the same dust attenuates all three lines, the line ratios fall along one \Aha value on an attenuation curve. While we only directly measure \pabeta/\halpha in MegaScience, we leverage the MOSDEF comparison sample to obtain an analogous measurement of the \halpha/\hbeta ratio (Balmer decrement), allowing us to constrain the attenuation curve. Since the MOSDEF comparison sample covers a very similar range of stellar mass, SFR, \Aha, and redshift as our MegaScience data (see Section \ref{subsec:mosdef_sample}), galaxies in both samples should have similar properties. Therefore, we would expect both samples to have the same average nebular attenuation curve. In particular, both \pabeta/\halpha and \halpha/\hbeta should fall along the same attenuation curve --- if they do not, then the assumed attenuation curve is likely incorrect. 

\begin{figure}[tp]
\vglue -5pt
\includegraphics[width=0.5\textwidth]{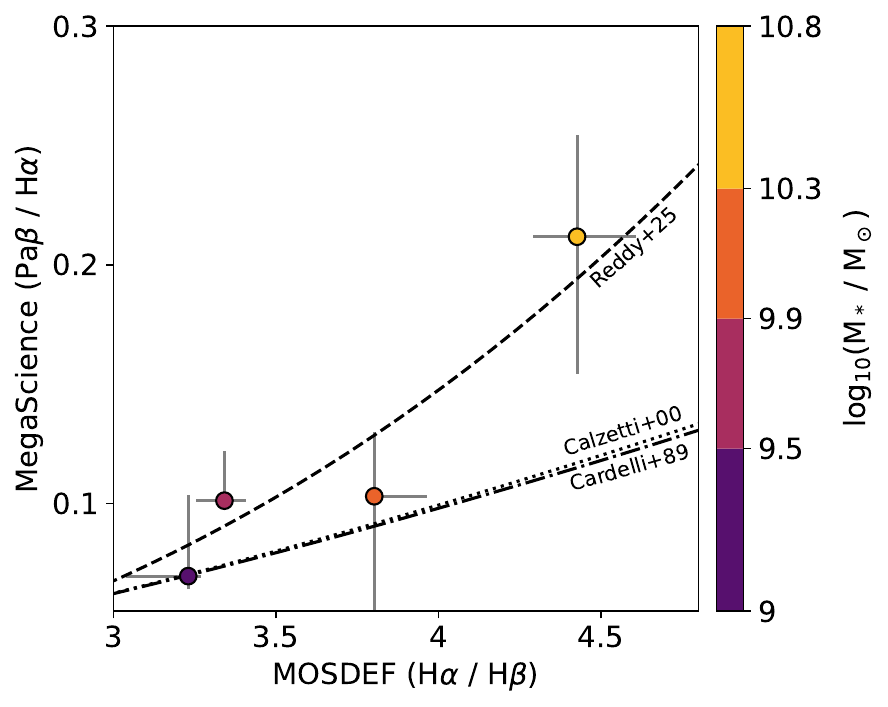}
\caption{
MegaScience \pabeta/\halpha line ratio vs. the \halpha/\hbeta for a comparable sample of MOSDEF galaxies (see Section \ref{subsec:mosdef_sample}). We present median line ratios from four mass bins that contain roughly equal numbers of MOSDEF galaxies, and their uncertainties are computed from bootstrapping. Predicted line ratios from the \citet{reddy_jwstaurora_2025} and \citet{cardelli_relationship_1989} nebular attenuation curves are shown. The data are consistent with the \citet{reddy_jwstaurora_2025} attenuation curve, particularly at high stellar masses. This consistency suggests that a shallower slope than \citet{cardelli_relationship_1989} is required at this mass and redshift regime.}
\label{fig:attenuation_curve}
\end{figure}

We aim to sample different inferred \Aha values along the attenuation curve. Since stellar mass correlates with \Aha, we bin the galaxies by the same stellar mass limits. We make four bins in $\log_{10}(M_*/M_\odot)$ that roughly contain the same number of MOSDEF galaxies: 9-9.5, 9.5-9.9, 9.9-10.3, 10.3-10.8. Within each bin, we measure the median \halpha/\hbeta from MOSDEF and the median \pabeta/\halpha from MegaScience, with uncertainties measured through bootstrapping. We note that we exclude MegaScience galaxies below $\log_{10}(M_*/M_\odot) = 9$, since the MOSDEF sample does not include dust measurements at such low masses. Our median measurements include upper limits on \pabeta/\halpha from MegaScience; while our MOSDEF \halpha/\hbeta measurements do require SNR$>3$ for both lines, the spectroscopic upper limits are considerably deeper than our photometric MegaScience measurements.

\begin{figure*}[tp]
\vglue -5pt
\includegraphics[width=\textwidth]{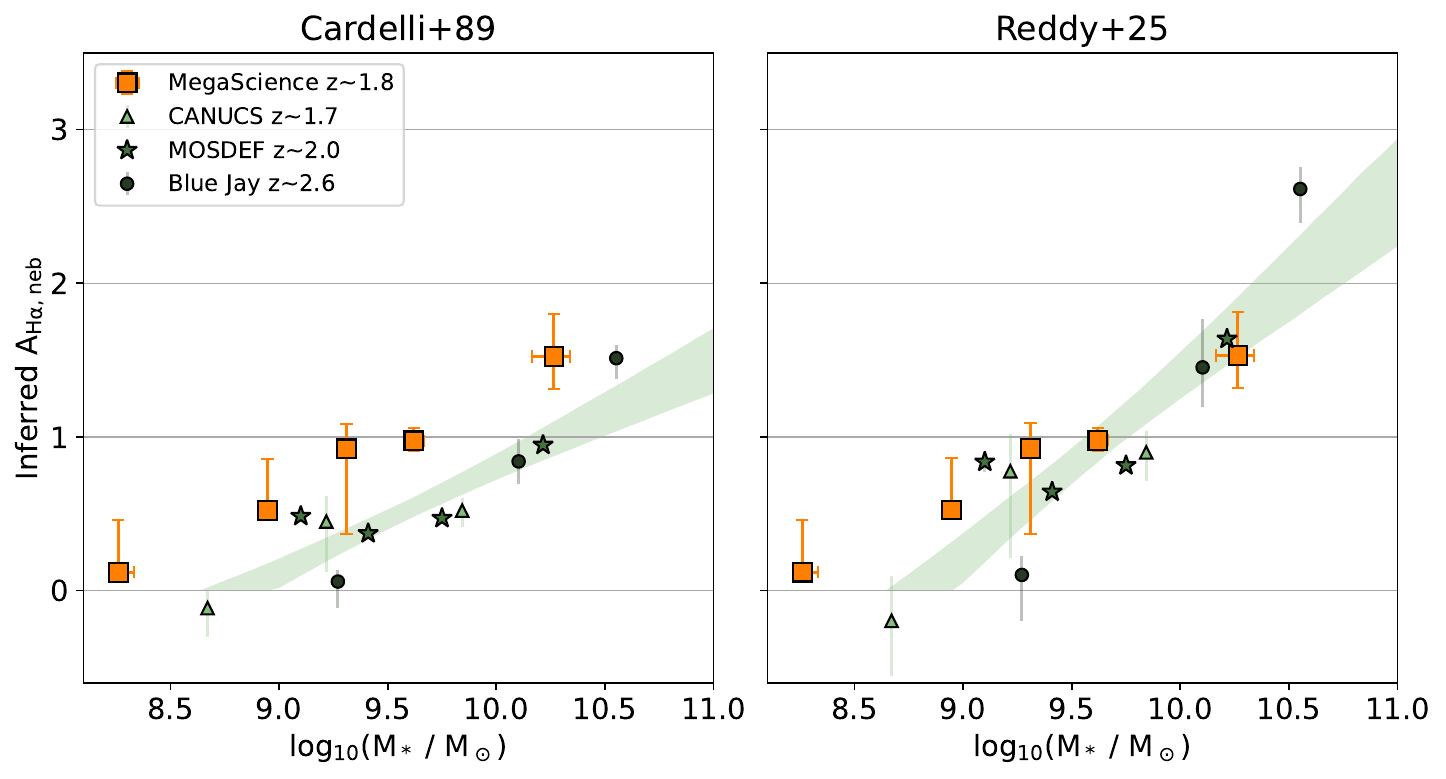}
\caption{
Median inferred \Ahaneb values for MegaScience (orange) compared to similar studies of star-forming galaxies at cosmic noon (CANUCS, MOSDEF, and Blue Jay). All other works use \halpha/\hbeta line ratios, while MegaScience uses the redder \pabeta/\halpha line ratio. The \Ahaneb are calculated assuming the \citet{cardelli_relationship_1989} attenuation curve (left) and the \citet{reddy_jwstaurora_2025} attenuation curve (right). The green shaded region shows the 1$\sigma$ fit to the comparison samples. We see that the \citet{cardelli_relationship_1989} attenuation curve causes our MegaScience measurements to be biased systematically high (median offset: 0.50mag), while the \citet{reddy_jwstaurora_2025} has a smaller median offset (0.34mag), especially at higher masses. We note that the different attenuation curves can cause wide variations in inferred \Ahaneb for the Balmer decrement samples, with over 1 magnitude of difference in the highest mass Blue Jay bin.}
\label{fig:comparison_to_others}
\end{figure*}

In Figure \ref{fig:attenuation_curve}, we plot our measured median \pabeta/\halpha line ratios against the MOSDEF median \halpha/\hbeta line ratios. We show the predicted line ratios in the presence of dust for the \citet{cardelli_relationship_1989}, \citet{calzetti_dust_2000}, and \citet{reddy_jwstaurora_2025} nebular attenuation curves. Given the MOSDEF Balmer decrements, the \pabeta/\halpha from MegaScience is much higher than predicted by the \citet{cardelli_relationship_1989} at the highest mass bin. At lower masses, the data are consistent with both curves. It appears that the shallower \citet{reddy_jwstaurora_2025} attenuation curve is more appropriate than the typical \citet{cardelli_relationship_1989} for these high mass star-forming galaxies at $z\sim2$. 

We further evaluate the attenuation curve through comparison to other similar studies. A variety of surveys with comparable data quality and methods have measured the Balmer decrement at cosmic noon, including MOSDEF \citep{kriek_mosfire_2015, shapley_first_2020}, CANUCS \citep{willott_canucs_2023, matharu_first_2023}, and Blue Jay \citep[GO 1810; PI Belli,][]{maheson_big_2025}. For each of these surveys, we bin Balmer decrement measurements by mass, then calculate the inferred \Ahaneb using both the \citet{cardelli_relationship_1989} and \citet{reddy_jwstaurora_2025} attenuation curves. We show these values alongside the \Ahaneb implied by our MegaScience \pabeta/\halpha line ratios in Figure \ref{fig:comparison_to_others}. Even though it has higher median redshift, we show the full Blue Jay sample in this figure. We verified that if we redshift-match the Blue Jay data to the MegaScience sample, there is very little change in their median mass and \Ahaneb.  

We use Figure \ref{fig:comparison_to_others} to evaluate how well our data based on \pabeta and \halpha agree with the \Ahaneb from prior observations based on Balmer lines. Since star-forming galaxies at cosmic noon should have roughly similar properties, we would expect all of the samples to fall on the same relation between \Ahaneb and mass, with some scatter. We linearly fit the data from the comparison Balmer line surveys (green curve in Figure~\ref{fig:comparison_to_others}) and use this fit to predict \Ahaneb in each bin. Then, we compute a median offset between our MegaScience measurements and these predicted \Ahaneb values. With the \citet{cardelli_relationship_1989} attenuation curve, the MegaScience measurements are systematically higher with a median offset among the four highest-mass bins of 0.47mag (left panel, Figure~\ref{fig:comparison_to_others}). The \citet{reddy_jwstaurora_2025} attenuation curve (right panel, Figure~\ref{fig:comparison_to_others}) gives a smaller median offset of 0.18mag, dropping to no offset at the high mass end. Thus, for low-mass galaxies, it is difficult to discriminate between the curves. However, for the high-mass galaxies, the \citet{reddy_jwstaurora_2025} curve is clearly a better fit.

We highlight that the choice of attenuation curve has large implications for the inferred \Ahaneb. In Figure \ref{fig:comparison_to_others}, we see that assuming the wrong attenuation curve can cause wide variations in \Ahaneb from the Balmer decrement. The median Blue Jay \Ahaneb measured with the \citet{cardelli_relationship_1989} attenuation curve is $\Ahaneb=0.84$mag, while \citet{reddy_jwstaurora_2025} gives median $\Ahaneb=1.45$mag. The discrepancy is even more dramatic for shorter-wavelength line ratios --- the maximum \Ahaneb from Blue Jay is 1.51mag with \citet{cardelli_relationship_1989}, but reaches 2.61mag with \citet{reddy_jwstaurora_2025}, more than 1 magnitude of difference. This offset could affect \halpha-derived SFRs by more than factor of 2. Therefore, it is especially important to constrain the attenuation curve when measuring dusty, high-mass galaxies.

\subsection{Axis Ratios}
\label{subsec:axis_ratio}

Dust geometry is closely linked with the relation between observed attenuation and viewing angle \citep{zuckerman_reproducing_2021, yuan_asymmetry_2021, giessen_probing_2022, lorenz_updated_2023, zhang_dust_2023}. In particular, if attenuating dust is spread throughout the interstellar medium (ISM), then we would expect \Aha to increase with decreasing axis ratio due to longer line of sight through edge-on galaxies.  With \Ahaneb measurements for  the MegaScience sample of galaxies assuming the \citet{reddy_jwstaurora_2025} attenuation curve, we now assess the effects of galaxy inclination on observed \Ahaneb. We divide the F150W axis ratio measurements into five bins with equal numbers of galaxies from the \halpha+\pabeta sample. In F150W, we have axis ratio measurements for 197 of the \ncombined objects in our sample, and 62 of the \ngals galaxies with both \halpha and \pabeta detections. 

\begin{figure}[tp]
\vglue -5pt
\includegraphics[width=0.5\textwidth]{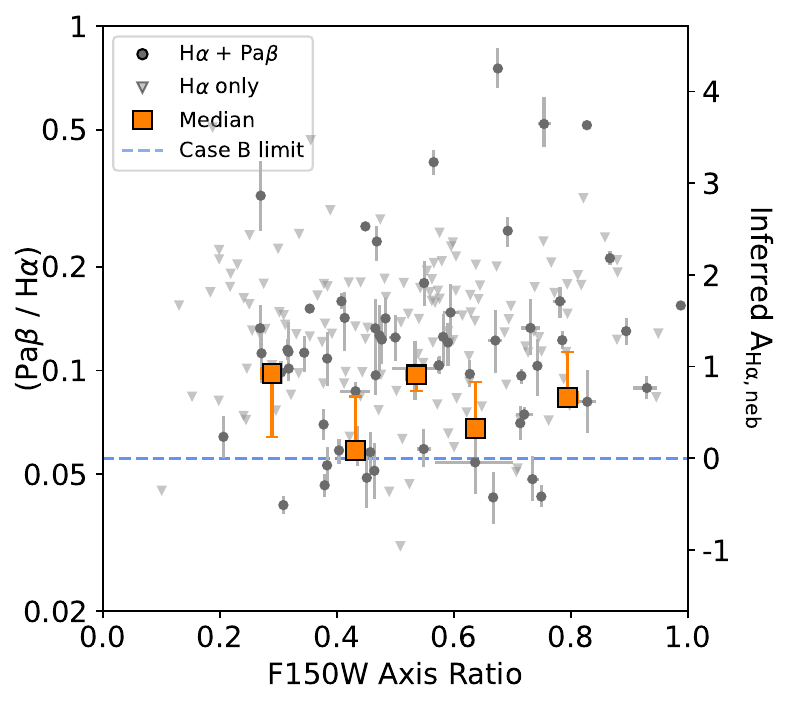}
\caption{Measured \pabeta/\halpha line ratio vs axis ratio, as determined by \texttt{Pysersic} fits to the photometry. We do not find a trend between the line ratio and axis ratio as measured in F150W (shown), F444W, or the medium band containing \halpha. Medians including the upper limits are shown in orange, with uncertainties measured through bootstrapping. The absence of any trend between dust content and inclination indicates that edge-on galaxies show similar nebular dust attenuation to face-on galaxies.}
\label{fig:dust_axisratio}
\end{figure}

In Figure \ref{fig:dust_axisratio} we show the measured \pabeta/\halpha vs. F150W axis ratio, with medians computed with the Turnbull estimator including the upper limits. There does not appear to be a trend with axis ratio, with all bins showing similar median line ratios. Thus, edge-on galaxies exhibit the same nebular attenuation values as face-on galaxies. While this figure shows F150W, we also compared against F444W axis ratio and against axis ratio in the medium band that contained \halpha, which included all \ncombined sources. In all cases, we found no trend between \Ahaneb and axis ratio. We also confirmed that there were no strong trends between the lensing magnification and axis ratio, as most of these galaxies are at similar magnification.

\section{Discussion}
\label{sec:Discussion}


The observed relations with mass, SFR, and axis ratio, along with the shape of the nebular attenuation curve, all have implications for the dust geometry in galaxies at cosmic noon. We explore what each of these results suggest individually, then discuss the possible geometries that can explain all of our observations. 

First, we find a strong trend of increasing \Ahaneb with stellar mass and SFR. Thus, more massive galaxies have higher dust content, which is expected given that the dust column density depends on the total amount of mass formed and the size of the attenuating regions \citep{lorenz_updated_2023}. These measurements do not strongly constrain the dust geometry, but do inform the quantity of dust. 

Next, we find that our measurements, particularly for the massive galaxies, appear to be consistent with a shallower attenuation curve \citep[e.g.,][]{reddy_jwstaurora_2025}. Similar results were observed in \citet{shivaei_diversity_2025}, where they show that the attenuation curve has a shallower slope at higher \AV values. Different properties of the dust grains may alter the attenuation curve, though it is more likely that the shallow curves are due to geometry effects. Dust covering fractions less than 1 can cause shallower slopes \citep{narayanan_theory_2018, salim_dust_2020, reddy_jwstaurora_2025}. In this model, some star-forming regions are dust-obscured, while others have little or no obscuration. Unobscured star-forming regions emit very strongly in blue wavelengths. Therefore, dust-free star-forming regions dominate the attenuation curve at short wavelengths, including the bluer Balmer lines. As wavelength increases, dust attenuation is less effective, so the obscured star-forming regions contribute more light. This change of contributions of the unobscured and obscured components with increasing wavelength leads to a flatter attenuation curve. Thus, the attenuation curve has a shallower slope when the dust covering fraction is less than 1. These lower covering fractions have been observed in prior works. By looking at \halpha and \pabeta maps of MegaScience galaxies, \citet{lorenz_measuring_2025} observed offsets between the \halpha emission and dust, indicating that some star-forming regions are likely unobscured. Additionally, \citet{reddy_paschen-line_2023} found non-uniform reddening of Balmer and Paschen lines in the same galaxies. Overall, the lower dust covering fraction seems to be a plausible explanation for the shallower attenuation curve. 

Finally, we find that there is no correlation between \Ahaneb and axis ratio in F150W, F444W, or the \halpha line medium band. This result is expected, as a number of studies have shown that nebular attenuation does not correlate with axis ratio across a range of redshifts \citep{yip_extinction_2010, battisti_characterizing_2017, yuan_asymmetry_2021, lorenz_updated_2023, maheson_big_2025}, although there may be a mild trend at $z\sim0$ \citep{chevallard_insights_2013}. Intuitively, an edge-on galaxy might  be expected to undergo more dust attenuation, since light escaping the galaxy must travel through a longer distance of ISM. However, if the attenuating dust is not uniformly spread throughout the ISM, the trend may disappear. Therefore, the lack of correlation between \Ahaneb and axis ratio indicates that the dust attenuating nebular emission is likely local to star-forming regions. As long as the star-forming regions are sufficiently small or sufficiently few in number such that they do not significantly overlap along line of sight, a dust geometry without a strong ISM dust component explains our observations.

We find that it is challenging to distinguish which attenuation curve is more applicable for low-mass galaxies, as they are reasonably consistent across samples with all attenuation curves. However, the high mass galaxies are clearly better fit by a shallow curve, indicating a dust geometry that is consistent with the multi-component models described in \citet{reddy_mosdef_2015} and \citet{lorenz_stacking_2024}. These models include a patchy and clumpy strong dust component, with very weak ISM contributions. Additionally, the suggested geometry becomes more complex and patchy at higher masses, potentially including kiloparsec-scale star-forming clumps. To explain the trends seen in this work, increasing \Ahaneb with stellar mass is explained by increasing dust column density in the clumps as well as additional clumps at higher masses. The shallow attenuation curve in the more massive galaxies is caused by non-uniform dust coverage of star-forming regions. Finally, the weak ISM component implies that \Ahaneb is independent of viewing angle, consistent with the lack of trend between inferred \Ahaneb and axis ratio shown in Figure~\ref{fig:dust_axisratio}.

\section{Summary} \label{sec:summary}

We used medium-band photometry from the MegaScience survey to measure \halpha and \pabeta emission line fluxes and ratios for a sample of \ngals galaxies at $1.2<z<2.4$, as well as \halpha fluxes for an additional \nhaonly galaxies. These measurements cover stellar masses as low as $\log_{10}(M_*/M_\odot) = 7.85$. From our emission line measurements, we derived nebular \Aha and compared it to stellar mass, SFR, and axis ratio. We used a mass-, SFR-, and redshift-matched sample of MOSDEF galaxies to compare our \pabeta/\halpha measurements to typical \halpha/\hbeta line ratios and determine the attenuation curve that best explains our measurements. We show that the assumed attenuation curve can have large effects on the assumed \Ahaneb and consequently derived SFRs. We summarize our findings:

\begin{itemize}
    \item The \pabeta/\halpha line ratio correlates strongly with stellar mass and SFR, implying that more massive galaxies have higher dust attenuation.
    \item The \citet{reddy_jwstaurora_2025} nebular attenuation curve is most consistent with the measured \pabeta/\halpha and \halpha/\hbeta line ratios for the more massive galaxies. In particular, the \citet{cardelli_relationship_1989} attenuation curve is too steep to explain our measurements at high masses. It seems likely that the shallower attenuation curve results from non-uniform dust covering fractions in the massive galaxies.
    \item Applying the \citet{reddy_jwstaurora_2025} attenuation curve to the \ngals galaxies with \pabeta detections, we find that the \Ahaneb ranges from 0 to 4.58, with a median of \Ahaneb= 1.24. These are significantly higher \Ahaneb values than would be implied by the \citep{cardelli_relationship_1989} attenuation law, by up to a full magnitude at high mass. Changing the dust law could thus increase the inferred \halpha SFR by up to a factor of two.
    \item Nebular \Aha shows no correlation with axis ratio, indicating that edge-on galaxies and face-on galaxies show similar nebular attenuation. Therefore, nebular attenuation is likely to be localized to small regions (relative to the galaxy size) rather than spread throughout the ISM. 
    \item Overall, the results are consistent with a patchy dust geometry in the massive galaxies, where nebular attenuation is primarily occurring in dusty star-forming clumps, as in \citet{reddy_mosdef_2015} and \citet{lorenz_updated_2023}. However, not all star formation takes place in these dusty clumps, and the obscuration of star-forming regions is non-uniform. This geometry leads to a lower covering fraction and therefore a shallower attenuation curve.
\end{itemize} 

Medium bands remain a powerful tool for measuring emission lines, allowing us to assemble dust measurements for a wide variety of galaxies, pushing to lower masses than many previous works. While the UNCOVER/MegaScience field contained 70,000 objects, the upcoming MINERVA survey promises to measure a sample ten times larger with medium bands, covering four of the five CANDELS extragalactic fields. Through this survey, we can construct an even larger sample of \pabeta/\halpha measurements, potentially further constraining the slope of the nebular attenuation curve. In addition, a much larger sample of \Ahaneb measurements allows us to solidify the relationship of dust to mass, SFR, and axis ratio.

While we have gained new insights into the dust geometry at cosmic noon, some outstanding issues remain. The average \citep{reddy_jwstaurora_2025} attenuation curve seems to describe our measurements well, but we note that \citep{reddy_jwstaurora_2025} find large variation in the attenuation curves for individual galaxies. Some of this variation may be driven by unique dust geometries or complex covering fractions, which will require more observations and modeling to fully understand. While the median measurements in this work follow the expected trends, there is large scatter in the individual measurements. It is unclear whether this scatter is due to measurement uncertainties or true variations between galaxies, such that each galaxy requires an individualized attenuation curve based on its dust geometry. Answering these questions in the near future is essential to accurately measure dust, mass, and SFR for galaxies at cosmic noon.

\begin{acknowledgments} 
This work is based on observations made with the NASA/ESA/CSA James Webb Space Telescope. The raw data were obtained from the Mikulski Archive for Space Telescopes at the Space Telescope Science Institute, which is operated by the Association of Universities for Research in Astronomy, Inc., under NASA contract NAS 5-03127 for \textit{JWST}. These observations are associated with JWST Cycle 1 GO program \#2651 and JWST Cycle 2 GO program \#4111, and this project has gratefully made use of a large number of public JWST programs in the Abell 2744 field including JWST-GO-2641, JWST-ERS-1324, JWST-DD-2756, JWST-GO-2883, JWST-GO-3538, and JWST-GO-3516. Support for program JWST-GO-4111 was provided by NASA through a grant from the Space Telescope Science Institute, which is operated by the Associations of Universities for Research in Astronomy, Incorporated, under NASA contract NAS5-26555.
\end{acknowledgments}

\vspace{5mm}
The JWST data presented in this article were obtained from the Mikulski Archive for Space Telescopes (MAST) at the Space Telescope Science Institute. The specific observations analyzed can be accessed via \dataset[doi: 10.17909/xgyb-9413]{https://doi.org/10.17909/xgyb-9413} and \dataset[doi: 10.17909/ebtc-tn86]{https://doi.org/10.17909/ebtc-tn86}.

\typeout{}\bibliography{MegaScience}{}
\bibliographystyle{aasjournal}

\end{document}